\documentclass[twocolumn,showpacs,showkeys,preprintnumbers,amsmath,amssymb,floatfix,prl]{revtex4-1}

\usepackage{graphicx}
\usepackage{bm}

\usepackage{placeins}
\usepackage{graphics}
\usepackage{color}
\usepackage{hyperref}

\begin{document}

\preprint{APS/123-QED}

\title{Proximity effects in bilayer graphene on monolayer WSe$_2$: Field-effect spin-valley
locking, spin-orbit valve, and spin transistor}

\author{Martin Gmitra and Jaroslav Fabian}

\affiliation{Institute for Theoretical Physics, University of Regensburg, 93040 Regensburg, Germany\\
}

\begin{abstract}
Proximity orbital and spin-orbit effects of bilayer graphene on 
monolayer WSe$_2$ are investigated from first-principles. We find that the
built-in electric field induces an orbital band gap of about 10 meV in bilayer graphene. Remarkably, 
the proximity spin-orbit splitting for holes is two orders of magnitude---the spin-orbit
splitting of the valence band at K is about 2 meV---more than for electrons. Effectively, 
holes experience spin-valley locking due to the strong proximity of the lower graphene
layer to WSe$_2$. However, applying an external transverse electric field of some 1 V/nm,
countering the built-in field of the heterostructure, completely reverses this effect and allows, instead
for holes, electrons to be spin-valley locked with 2 meV spin-orbit splitting. Such a behavior constitutes a highly efficient field-effect spin-orbit
valve, making bilayer graphene on WSe$_2$ a potential platform for a field-effect
spin transistor.  
\end{abstract}

\maketitle

\paragraph{Introduction.}
Heterostructures of two-dimensional materials can
fundamentally alter their properties due to proximity effects.
For example, graphene on transition metal dichalcogenides (TMDC)
can serve as a new platform for optospintronics~\cite{Gmitra2015:PRB},
as recently also demonstrated experimentally~\cite{Avsar2017:arXiv,Luo2017:AcsNL}
promoting graphene spintronics~\cite{Han2014:NatNano} 
towards applications~\cite{Zutic2004:RMP, Fabian2007:APS}.
Bilayer graphene (BLG) on TMDC is expected to represent
even more technologically feasible approach as it allows a precise (sub meV) control
of the chemical potential---due to much smaller Fermi level
fluctuations~\cite{Rutter2011:NP}---than in single layer graphene~\cite{Martin2008:NP}.

There have recently been intensive efforts to predict realistic graphene structures, 
through enhancing spin-orbit coupling by decorating graphene with adatoms, that would exhibit
quantum spin (and anomalous) Hall effects~\cite{Qiao2010:PRB, Weeks2011:PRX, Zhang2012:PRL, qiao_quantum_2014}, 
introduced by Kane and Mele~\cite{Kane2005:PRL} as a precursor of topological 
insulators~\cite{bernevig_quantum_2006, konig_quantum_2007, zhang_topological_2009}.
Unlike promising approaches to enhance spin-orbit coupling via
adatoms~\cite{Neto2009:PRL,Gmitra2013:PRL}, demonstrated already
experimentally by the giant spin Hall effect signals~\cite{balakrishnan_colossal_2013, Avsar2014:NatComm}, 
van der Waals heterostructures provide more robust control 
towards technological reproducibility of devices.
Recently, proximity effects in graphene on the whole family of TMDCs 
as potential substrates for graphene were explored theoretically~\cite{Kaloni2014:APL,Gmitra2016:PRB}.
An enhancement of proximity spin-orbit coupling, of about 1 meV,
was predicted, which is giant compared to bare graphene, in which spin-orbit coupling
is about 10 $\mu$eV~\cite{Gmitra2009:PRB}.
The special case is graphene on WSe$_2$, where the predicted band inversion was proposed
to lead to novel topological properties,
\cite{Gmitra2016:PRB,Morpurgo2015:NComm,Yang2016:2DM},
and giant spin relaxation anisotropy~\cite{Cummings2017:arXiv}.
Important, graphene/TMDCs has already been grown~\cite{Lin2014:ACS,Lin2014:APL,Ugeda2014:NatMat,Azizi2015:ACS}
and investigated for transport~\cite{Lu2014:PRL,Larentis2014:NL,Avsar2014:NatComm,Morpurgo2015:NComm,Omar2017:PRB},
optoelectronics~\cite{Massicotte2016:NatNano}
as well as considered for technological applications~\cite{Bertolazzi2013:ACSNano,Roy2013:NatNanotech,Zhang2014:SREP,Kumar2015:MT}.

A BLG can exhibit an electronic bandgap in the presence of a transverse 
electric field~\cite{Ohta2006:Science,Oostinga2008:NM,Zhang2009:N,McCann2006:PRL}.
The tunable bandgap enables a variety of different device concepts with 
novel functionalities for electronic, optoelectronic, and sensor applications. 
There were several proposals to increase the ON/OFF ratio in gated BLG, 
introducing a tunnel field-effect transistor~\cite{Alymov2016:SciRep}
or  a field effect transistor by adsorbate doping~\cite{Szafranek2011:NL}
to establish a displacement field. Also, BLG/TMDC heterostructures can 
potentially realize predicted topological insulating phases protected by 
no-valley mixing symmetry, featuring quantum 
valley Hall effects and chiral edge states~\cite{Wang2009:PRL,Zhang2011:PRL,Ju2015:N}.

\begin{figure}[h!]
	\includegraphics[width=0.99\columnwidth]{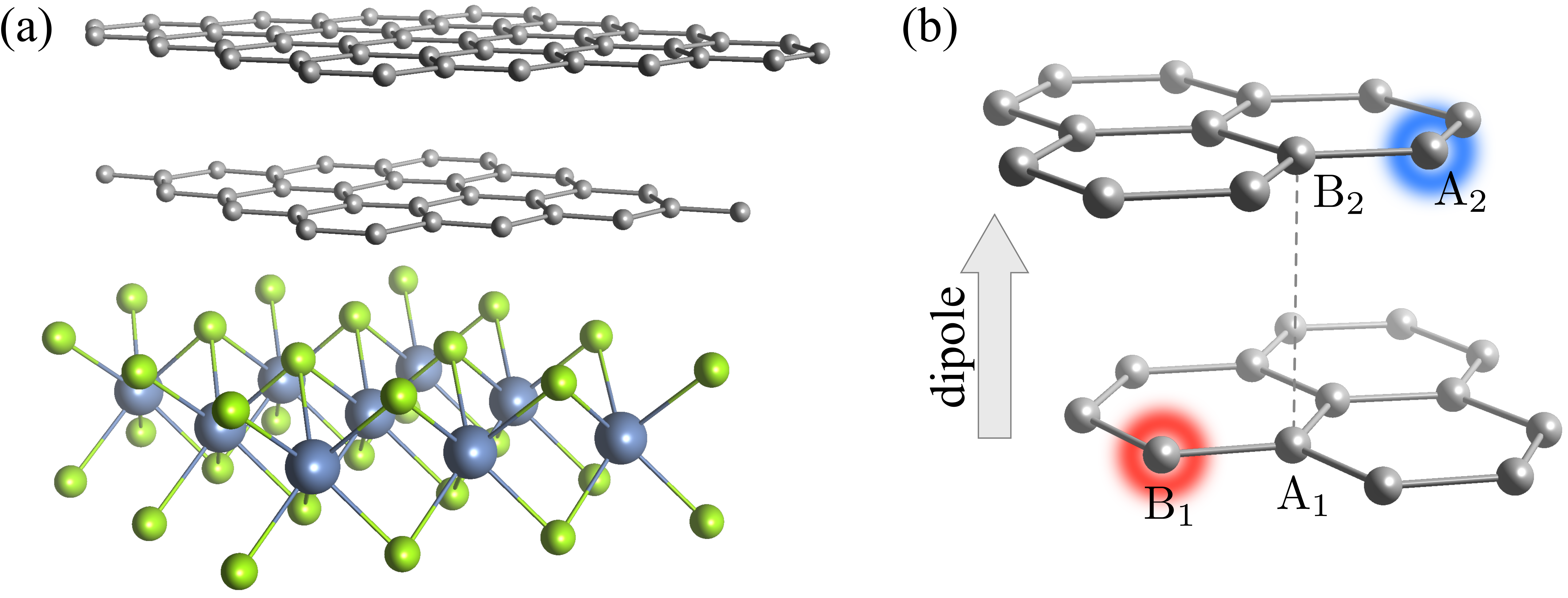}
	\caption{(a) Atomic structure of
		bilayer graphene on a monolayer WSe$_2$ (supercell). (b)
		Sketch of bilayer graphene with atom labels forming a bare
		unit cell in Bernal stacking. Orbitals on non-dimer atoms B$_1$ and A$_2$
		form the low energy valence and conduction bands in the electronic structure 
		of bilayer graphene, with B$_1$ being closer to WSe$_2$.}
	\label{Fig:scheme}
\end{figure}

In this paper we find, by performing first-principles investigations, 
that in a BLG/WSe$_2$ heterostructure a displacement field emerges intrinsically, allowing for a highly efficient electric control of proximity effects. 
Specifically, we find that
(i)~The intrinsic bandgap, which is about 10 meV, can be enhanced, reduced to zero,
or reversed by typical experimental electric fields on the order of 1 V/nm;  
(ii)~The spin-orbit coupling of the valence band is giant,
about 2 meV, being two orders of magnitude greater than
in the conduction band, which is similar to intrinsic BLG 
\cite{Konschuh2012:PRB}. The reason for this huge disparity
is that the valence band is formed by non-dimer carbon 
atom orbitals in the bottom layer adjacent to WSe$_2$, while the
conduction band is formed by non-dimer orbitals in the
top layer, where proximity effects are naturally weak
(the pair of atoms vertically connected we call dimer, the other pair non-dimer);
(iii) The spin-orbit coupling of the valence bands is of
spin-valley locking character, inherited from the monolayer WSe$_2$ substrate;
(iv)~A transverse electric field can turn spin-orbit coupling and spin-valley locking 
of electrons effectively on (and holes off), by countering the built-in
field. We call this effect {\it spin-orbit valve}. Connecting the
spin-orbit to spin relaxation, the two decades of spin-orbit coupling
translates into four orders of magnitude change in spin relaxation. 
Such a strong field-effect spin relaxation effect would be an ideal
platform for the spin transistor of Hall and Flatte~\cite{Hall2006:APL}.

\begin{figure}[h!]
 \includegraphics[width=0.99\columnwidth]{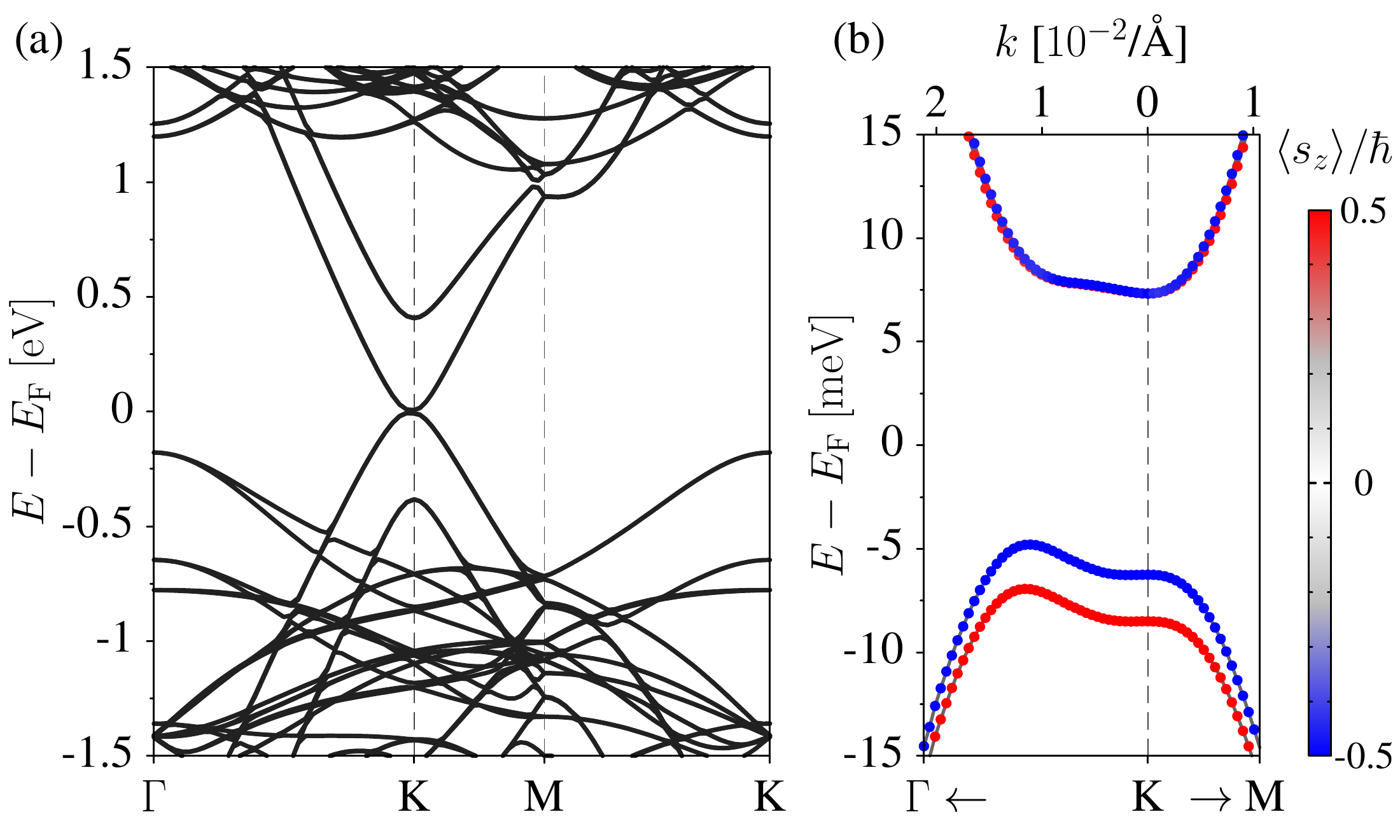}
 \caption{(Color online) (a) Calculated electronic band structure of bilayer graphene
 on monolayer WSe$_2$. (b)~Zoom to the fine structure of the low energy bands 
 close to the Fermi level. Bands with positive (negative) $z$ component of the spin 
 are shown in red (blue).
 }\label{Fig:bands}
\end{figure}

\paragraph{Electronic band structure of bilayer graphene on WSe$_2$.}
The electronic structure calculations and structural relaxation were performed
by {Quantum ESPRESSO}\cite{Giannozzi2009:JPCM}; see Supplemental material~\cite{SM} 
for further details~\cite{Hohenberg1964:PR,Perdew1996:PRL,Grimme2006:JCC,Barone2009:JCC,Bengtsson1999:PRB}.
In Fig.~\ref{Fig:bands}(a) we show the calculated electronic band structure
of BLG on monolayer WSe$_2$ along high symmetry lines. 
The parabolic band dispersion of high  and low energy bands 
close to the Fermi level resembles bare BLG~\cite{Nilsson2008:PRB,McCann2006:PRL}.
The high energy bands originate from the orbitals in dimer A$_1$ and
B$_2$ atoms connected by direct interlayer hopping~\cite{Nilsson2008:PRB,McCann2006:PRL}
which shifts the bands some 400~meV off the Fermi level, far enough to 
ignore these bands for transport.

The proximity effects influence mainly the low energy bands of BLG.
These bands originate from the $p_z$ orbitals on non-dimer B$_1$
and A$_2$ atoms which form the valence and conduction band edges, respectively.
An indirect bandgap of 12~meV, see Fig.~\ref{Fig:bands}(b),
is opened due to proximity induced intrinsic electric field built 
across the BLG/WSe$_2$ heterostructure. The transverse field points 
from WSe$_2$ towards the BLG (we call this direction positive) 
with the amplitude of 0.267~V/nm. This is why B$_1$  electrons
have lower energy, and form the valence band, while A$_2$ electrons
have higher energy and form the conduction band. 
Apart from the orbital effect, the proximity
also induces significant spin splitting of 2.2~meV in the valence band, 
seen in Fig.~\ref{Fig:bands}(b).
This makes sense as B$_1$ atoms, responsible for the valence band, 
are close to WSe$_2$ and experience the proximity effects most. 

Proximity effects in the conduction band are essentially
non-existent, since A$_2$ atoms sit far from WSe$_2$. The numerical
value of the spin-orbit splitting for the conduction band obtained by
Quantum ESPRESSO is too small (about 3 $\mu$eV), as $d$ orbitals are not properly
treated by the method. We know that in pristine BLG spin-orbit splitting should be
about 24 $\mu$eV, due to the presence of $d$ orbitals 
\cite{Konschuh2012:PRB}. We can safely assume that this value,
(perhaps up to 10\% higher or lower due to the proximity effects
on $p$ orbitals), of spin-orbit splitting is there for BLG on WSe$_2$.
We conclude that holes experience spin-orbit coupling (2 meV)
 two orders of magnitude higher than electrons (20 $\mu$eV).
 A recent experiment has observed Shubnikov-de Haas oscillations in BLG on 
WSe$_2$ and found, by fitting the results to a simple band structure model, 
that spin-orbit proximity effect is about 10 meV \cite{Morpurgo2016:PRX}. 
Our results presented here disagree with this interpretation.

\begin{figure}[h!]
 \includegraphics[width=0.99\columnwidth]{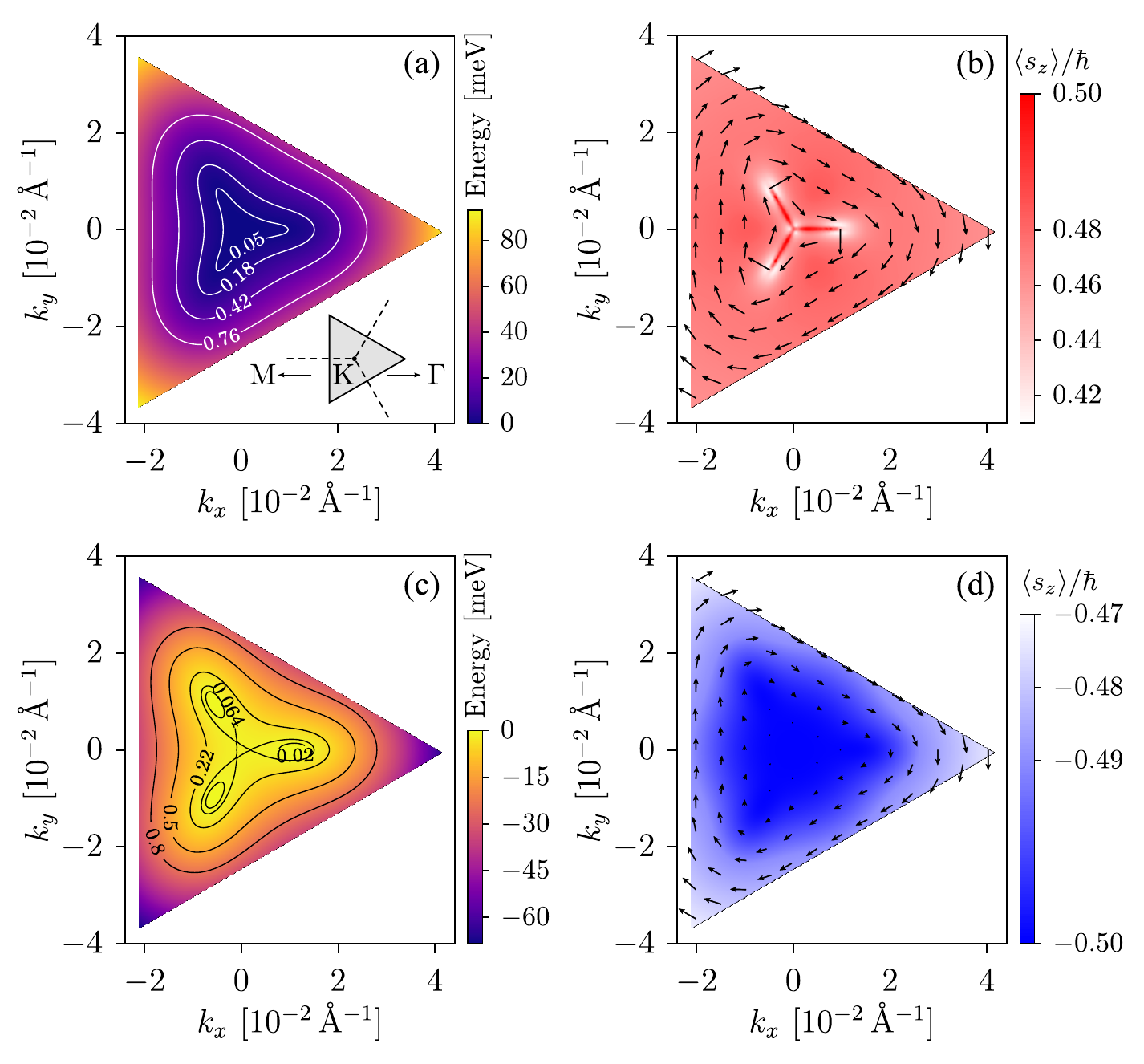}
 \caption{(Color online) Calculated low energy electronic and spin properties
 of bilayer graphene on a monolayer WSe$_2$. Shown are color map plots centered at
 K point representing 0.5\% of the first Brillouin zone area for
 (a)~energy of bottom conduction band measured from the conduction band edge. 
 The contours correspond to carrier concentrations for 0.05, 0.18, 0.42 
 and 0.76~$\times 10^{12}$~cm$^{-2}$. The inset depicts cut of the first 
 Brillouin zone, shown by dashed lines, near the K point with directions 
 towards M and $\Gamma$ points.
 (b)~Color map of $z$ component spin expectation value with in-plane spin 
 textures shown by the arrows.
 (c)~Energy of the valence band measured from the valence band edge 
 with contours showing carrier concentration of 0.02, 0.064, 0.22, 0.5 and 
 0.8$\times 10^{12}$~cm$^{-2}$, and
 (d)~similar as in (b) but for the valence band.
 }\label{Fig:maps}
\end{figure}

\paragraph{Valley spin-orbital effects.}
Inspecting energy dispersions near the K valley for the spin split
low energy bands, we find pronounced trigonal warping, see color map 
plots in the $k_x$ and $k_y$ momentum plane in Fig.~\ref{Fig:maps}(a,c).
Only the bottom spin-orbit split conduction band and the top valence bands are shown. 
The area of the plots corresponds to 0.5\% of the full first 
Brillouin zone. The calculated spin expectation values for 
the low energy states are principally locked out-of-plane,
see color map plots in Fig.~\ref{Fig:maps}(b,d).

For low carrier concentrations, the Fermi contours form three pockets
along K--$\Gamma$ directions. Further increase of doping level
merges the three Fermi pockets with an emerging pocket centered at
the K point. For the top valence band the merging occurs at the carrier
concentration of ${\rm 0.064 \times 10^{12}\,cm^{-2}}$. This is 
accompanied by the presence of a van Hove singularity in the 
density of states~\cite{SM}. The same holds for the spin-orbit split band.
The second van Hove singularity appears at the energy lower by 2.2~meV,
that corresponds to the proximity induced spin-orbit splitting.
A further increase of the Fermi level leads to a linear increase 
of the Fermi surface area and carrier concentrations~\cite{SM}.
Multiplying the calculated carrier concentration~\cite{SM} by $h/2e$ 
we can estimate the inverse frequency of the Shubnikov-de Haas
oscillations, which for carrier concentrations 
of about ${\rm 0.2 \times 10^{12}\,cm^{-2}}$ corresponds to 
${\rm 10~T^{-1}}$. The above mentioned experiment~\cite{Morpurgo2016:PRX}
observes similar values, although we cannot make a quantitative
comparison due to the absence of carrier density data.

\paragraph{Spin-orbit valve.}

\begin{figure*}[h]
 \includegraphics[width=1.80\columnwidth]{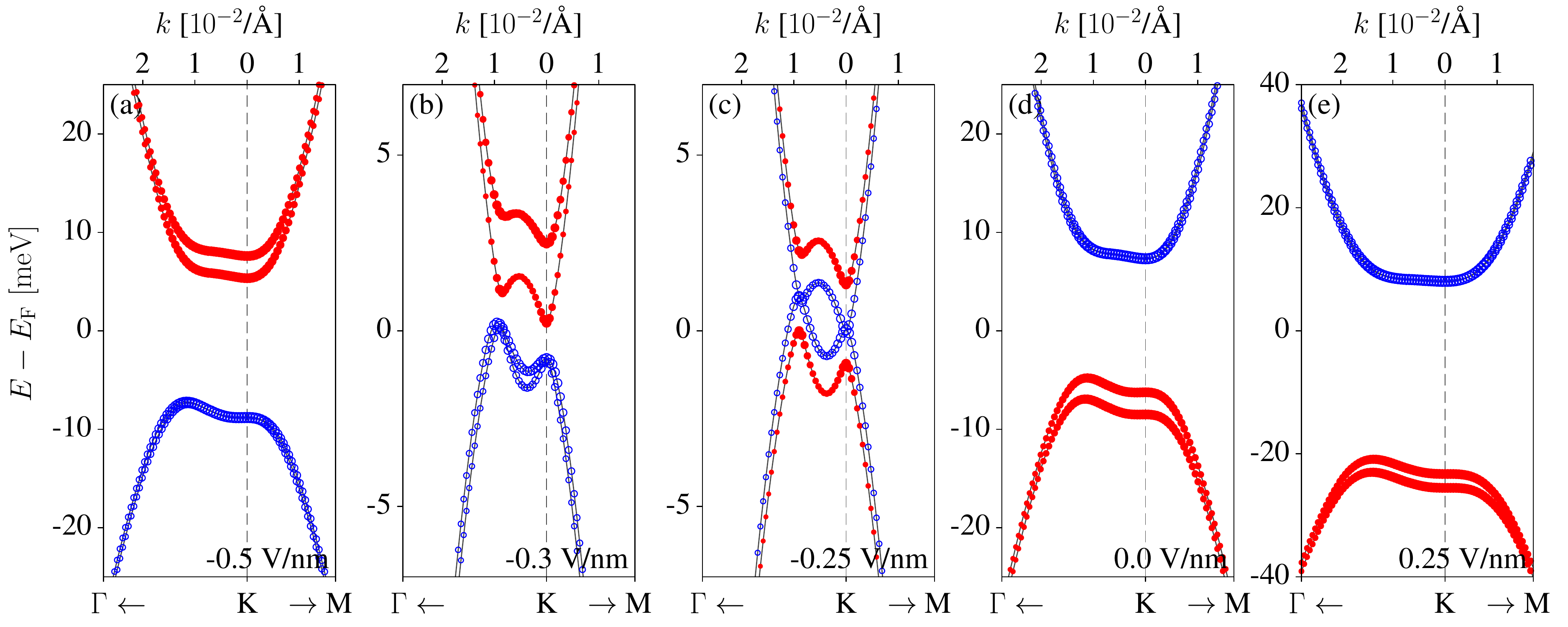}
 \caption{(Color online) Calculated sublattice resolved band structures 
 around K valley for transverse electric field of 
 (a)~{-0.5}~V/nm, (b)~{-0.2979}~V/nm, (c)~{-0.25}~V/nm, (d)~zero field, and (e)~0.25~V/nm.
 The circles radii correspond to the probability of the state 
 being localized on carbon atoms B$_1$ (red) filled circles and 
 atoms A$_2$ (blue) open circles.
 }\label{Fig:bands_chain}
\end{figure*}

All electrical control of  spin and orbital properties is a key
feature for spintronics devices.
Proximity induced spin-orbit coupling and the intrinsic electric polarization
in BLG/WSe$_2$ heterostructure can be efficiently controlled by 
an applied  transverse electric field, as shows the plot of low energy band 
structures of BLG in the presence of applied electric fields in
Fig.~\ref{Fig:bands_chain}.
Spontaneous polarization of the heterostructure induces
a dipole in the simulated cell of about 0.7~Debye. This gives rise to
a built-in transverse electric field of 0.267~V/nm which opens
the electronic bandgap in BLG to the value of 12~meV. 
Applying a positive electric field the electronic gap further opens
as the external field adds to the built-in internal field, 
see Fig.~\ref{Fig:bands_chain}(e). However, 
when the applied field direction is reversed, the band gap shrinks, 
and for the field of -0.25~V/nm which nearly compensates the intrinsic
field, the gap fully closes, see Fig.~\ref{Fig:bands_chain}(c). 
A further decrease (increase of the negative amplitude)
of the field the bandgap opens again, but the characters
of the valence and conduction bands flip and we get 
a spin-orbit valve!

The reason for this effect is simple. 
At zero applied electric field the spin-orbit splitting of the low energy
valence bands originates from the bottom layer of BLG, adjacent
to  WSe$_2$. More specifically, as already mentioned, the bands originate from the orbitals 
on carbon atoms B$_1$. On the contrary, the low energy conduction 
bands are localized on the top (remote) BLG layer, specifically
on atoms A$_2$, see Fig.~\ref{Fig:bands_chain}(d). The
spin-orbit splitting of the valence bands is about 100 times larger
near the K valley in comparison to the conduction band splitting.
In the built-in electric field the bottom BLG layer experiences
a lower potential then the top layer. Therefore, the valence
(occupied) states originate from the bottom BLG layer, and the
conduction (unoccupied) states from the top layer. For negative
applied field of -0.5~V/nm the potential across the BLG reverses
and band character switches, compare Fig.~\ref{Fig:bands_chain}(a) and (d).
Applying the external electric field induces also changes in 
the energy offset of the low energy bands with respect to the valence band
maximum of the WSe$_2$. For negative fields the BLG valence top is 
pushed down in energy and for the fields below -1~V/nm the valence
top of the WSe$_2$ is above the valence top of BLG. In effect, BLG gets
electron doped~\cite{SM}.

\paragraph{Bilayer graphene spin transistor.}

The proposed electrical switching of the spin-orbit splitting
(either by changing the doping between electrons and holes, 
or by changing the electric field at a fixed chemical potential)
presents a unique opportunity for a novel spin transistor design.
We build on the spin transistor proposed by Hall and Flatte~\cite{Hall2006:APL},
which is an alternative to charge-based transistors by overcoming
the $k_B T$ barrier for ON/OFF operations. 

\begin{figure}[h!]
\begin{tabular}{l}
(a)\\
 \includegraphics[width=0.99\columnwidth]{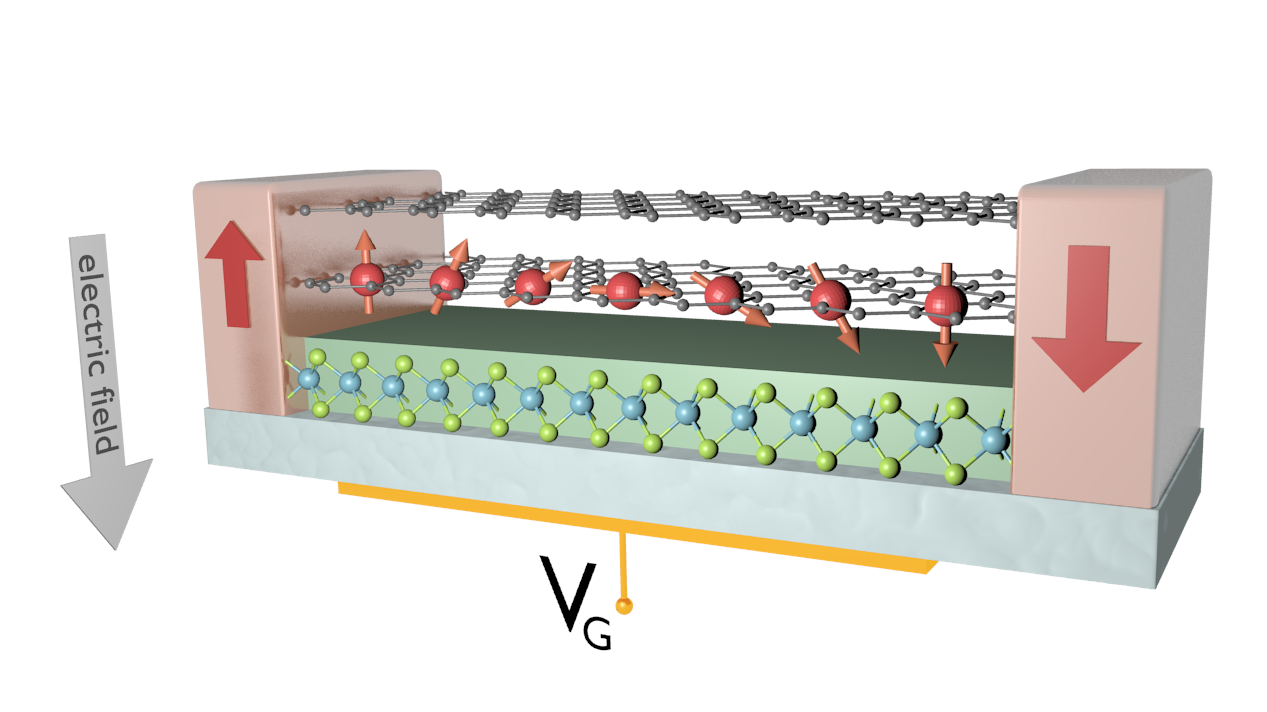}\\
(b)\\	
 \includegraphics[width=0.99\columnwidth]{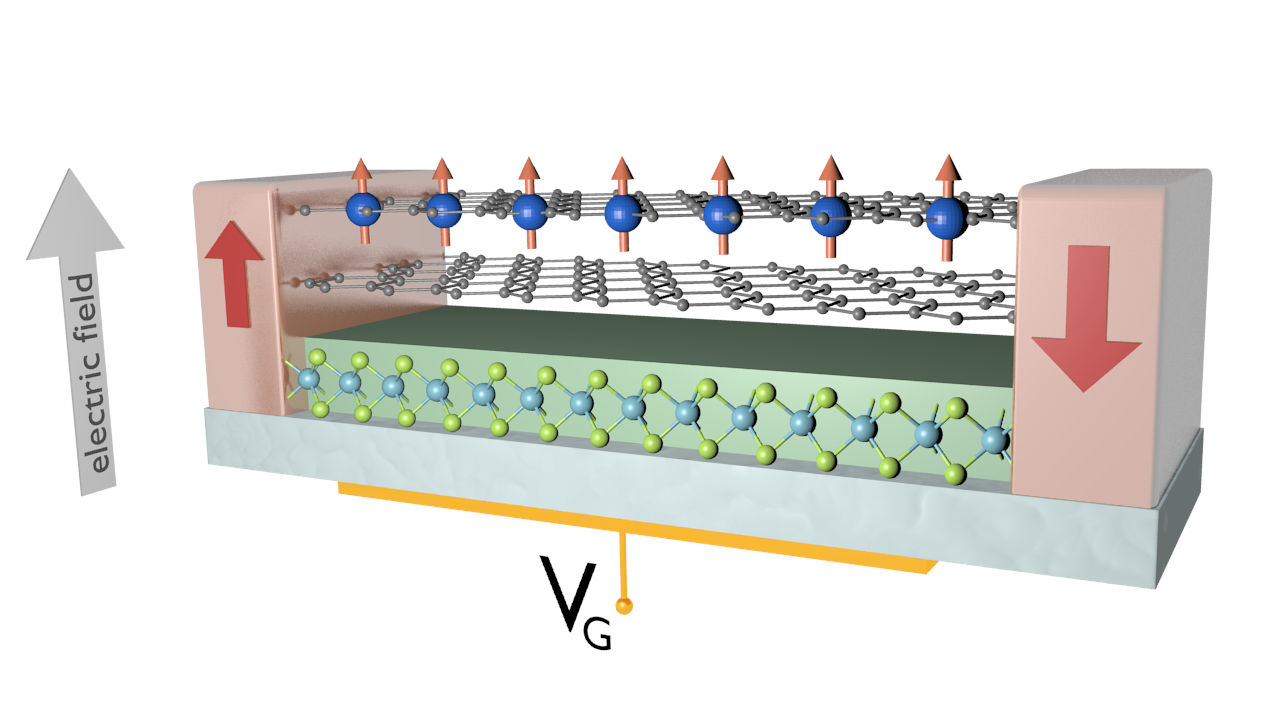}
\end{tabular}
 \caption{(Color online) Schematics of spin-field effect transistor
 	of bilayer graphene on a transition metal-dichalcogenide with
 	two ferromagnets in antiparallel configuration acting as injector and 
 	detector of spins
 	in the (a)~spin-ON and (b)~spin-OFF state. 
 }\label{Fig:transistor}
\end{figure}

Suppose we have a half-metallic spin injector and detector, in an antiparallel
configuration, connected to BLG proximitized to TMDC, see Fig.~\ref{Fig:transistor}. 
We can control the spin-orbit coupling of the carriers in BLG by the spin-orbit valve
effect. In the ON state, the spin-orbit coupling is high, spin relaxation
is fast, and spin in the channel is reduced. Large current flows.
In the OFF state, spin-orbit coupling is weak, spin relaxation slow, 
and spin in the channel is preserved. No current (in ideal case) flows. 

The reason why this works is that spin relaxation depends on the 
square of spin-orbit coupling. BLG on WSe$_2$ should have spin relaxation
due to the D'yakonov-Perel' \cite{Dyakonov1971}, 
which is as a motional 
narrowing of the spin precession in a fluctuating 
(due to momentum scattering) emerging spin-orbit
field $\mathbf{\Omega}(k)$. The spin-orbit splitting energy is proportional 
to $\hbar \Omega$, where $\Omega$ is the averaged 
spin-orbit field for a Fermi contour. The spin relaxation
rate is then given by $1/\tau_{\rm s}=\Omega^2 \tau$, where
$\tau$ is the momentum relaxation time. 
Having calculated spin splitting at zero field of 2.2~meV for 
the low energy bands, we estimate the spin relaxation time 
of 1~ps, assuming typical value for $\tau = 100$~fs. 
Applying electric field of -0.5~V/nm, the band character switches
and spin relaxation is reduced, which is comparable
to what is seen in ultraclean graphene~\cite{Guimaraes2014:PRL,Gurram2016:PRB} 
and BLG~\cite{Ingla2015:PRB} encapsulated in hBN.
The expected field-effect variation of spin relaxation time in BLG on WSe$_2$
is 4 orders of magnitude! Such a modulation is, to the best 
of our knowledge, unprecedented.

In conclusion, we have studied from first principles
the electronic structure of bilayer graphene on WSe$_2$. The most
important finding is the field-effect spin-orbit valve, 
allowing for an efficient switching, by two orders of magnitude, 
of spin-orbit coupling of electrons and holes.  

\begin{acknowledgments}
This work was supported by DFG SFB~689 and by 
the European Union’s Horizon 2020 research and
innovation programme under Grant agreement No. 696656.
The authors gratefully acknowledge the Gauss Center for
for Supercomputing e.V. for providing computational 
resources on the GCS Supercomputer SuperMUC at Leibniz
Supercomputing Center.
\end{acknowledgments}

\bibliography{paper}

\newpage
\section{Supplemental Material}
The supplemental material provides further details of the calculational
methods, calculated 
electronic properties, and electrostatic quantities. The information 
can be useful for further realization of proposed spin transistor
concept in transport experiments.

Electronic structure of bilayer graphene (BLG) on WSe$_2$ was calculated by 
means of density functional theory~\cite{Hohenberg1964:PR}.
We consider a supercell structural model, shown in Fig.~1(a) in the paper, containing a $3\times 3$ cell 
of WSe$_2$ and a $4\times 4$ cell of BLG in Bernal stacking, 
see Fig.~1(b) in the paper, with common lattice 
constant of 9.8934~\AA. The supercell has 91 atoms.
In such a quasi commensurate structure the residual strain 
results in a stretching of BLG by only 0.5\%.
Similar quasicommensurate superstructures of 
transition metal dichalcogenides 
have been grown on HOPG~\cite{Ugeda2014:NatMat}.
Electronic structure calculations and structural relaxation were performed
within plane wave package {Quantum ESPRESSO}~\cite{Giannozzi2009:JPCM},
using norm conserving pseudopotentials with kinetic energy cutoff 
of 60~Ry for wavefunctions. For the exchange-correlation potential 
we used the generalized gradient approximation~\cite{Perdew1996:PRL}.
The vacuum of 15~${\rm\AA}$ normal to the layer planes was considered. 
The first Brillouin zone was sampled with 545 $k$ points. 
We estimated error in underestimation of the electronic bandgap 
from the convergence of the number of the used $k$ points 
to 0.5~meV. We note that spin-orbit coupling properties 
for the used number of $k$ points are well converged. The atomic positions 
were relaxed using the quasi-Newton algorithm based on the trust radius 
procedure including the van der Waals interaction which was treated 
within a semiempirical approach~\cite{Grimme2006:JCC,Barone2009:JCC}.
The average interlayer distance 3.376~\AA~between BLG and WSe$_2$
is of van der Waals order. In our calculations we applied 
the dipole correction~\cite{Bengtsson1999:PRB}, which turned out 
to be crucial to get numerically accurate band offsets and
internal electric fields.

\begin{figure}[h!]
 \includegraphics[width=0.99\columnwidth]{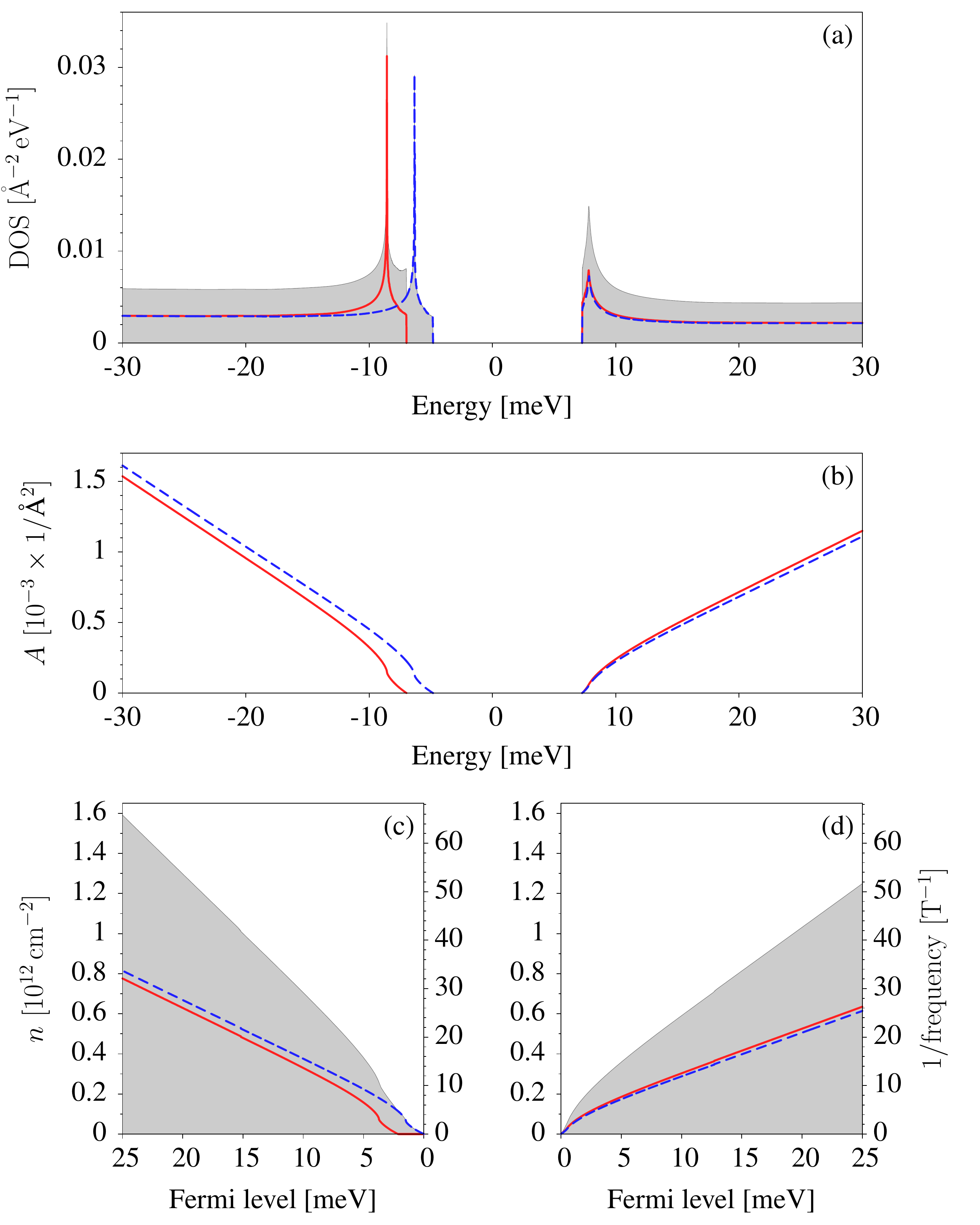}
 \caption{Calculated 
(a)~total density of states per unit volume per unit of energy, 
shown by the shaded area. Contributions from spin split bands are shown
by the lines.
(b)~Area of the Fermi contour band resolved as in (a).
(c)~Carrier concentration as a function of Fermi level for hole doping,
shown by the shaded region with the corresponding band contribution
shown by lines.
(d)~Carrier concentrations as in (c) but for electron doping.
 }\label{Fig:dos}
\end{figure}

In Fig.~\ref{Fig:dos}(a) we show calculated density of states for
BLG on WSe$_2$. Valence and conduction band edges 
are separated by the electronic gap of 12~meV. Near the edges
pronounced van Hove singularities are present. They appear 
in pairs from each subband separately. As the valence bands
are strongly spin split by 2.2~meV the peaks in the density
of states are much more pronounced. The splitting of the conduction 
bands near the band edge is reduced 800 times. We note
that the van Hove singularities appear at the energies at which the
trigonal symmetric Fermi pockets extending along K--$\Gamma$ path
near the band edges merge with the pocket centered at the K point.

Corresponding band resolved areas of the Fermi contours are shown
in Fig.~\ref{Fig:dos}(b). For energies about 5~meV from the band edges
the Fermi contour areas grow linearly. The nonlinearity near the
band edge is related with the existence of more then one Fermi pocket
within a subband. Calculated band resolved carrier concentration 
shown in Fig.~\ref{Fig:dos}(c) and (d), show similar dependence
on energy as the Fermi contours.

\begin{figure}
 \includegraphics[width=0.98\columnwidth]{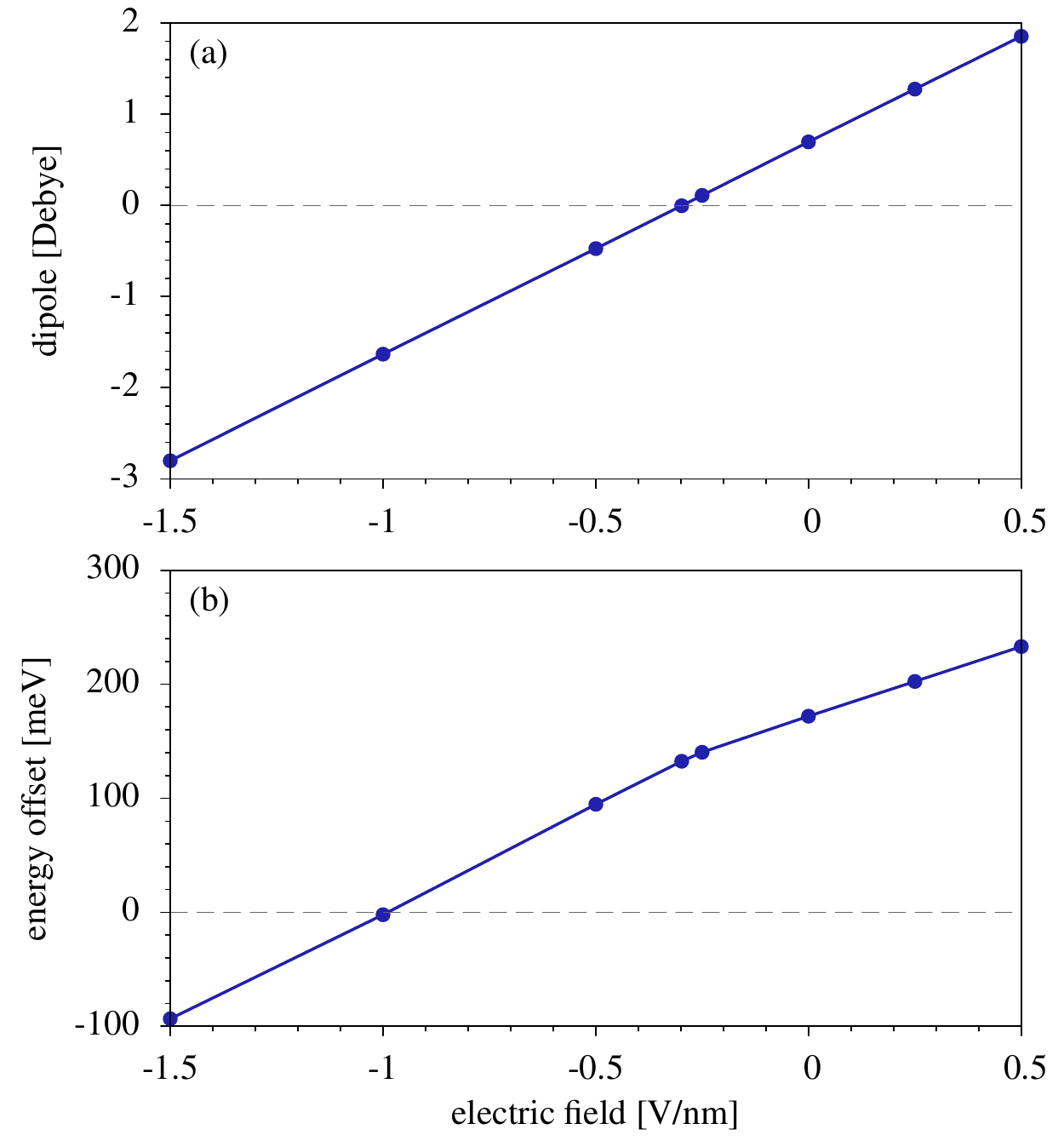}
 \caption{Calculated electric field dependences in bilayer graphene on
 WSe$_2$ of
(a)~dipole induced in the heterostructure, and
(b)~energy offset of the bilayer graphene valence band at K point and top of the valence band maximum of the WSe$_2$.
 }\label{Fig:dipole}
\end{figure}
\begin{figure}
	\includegraphics[width=0.98\columnwidth]{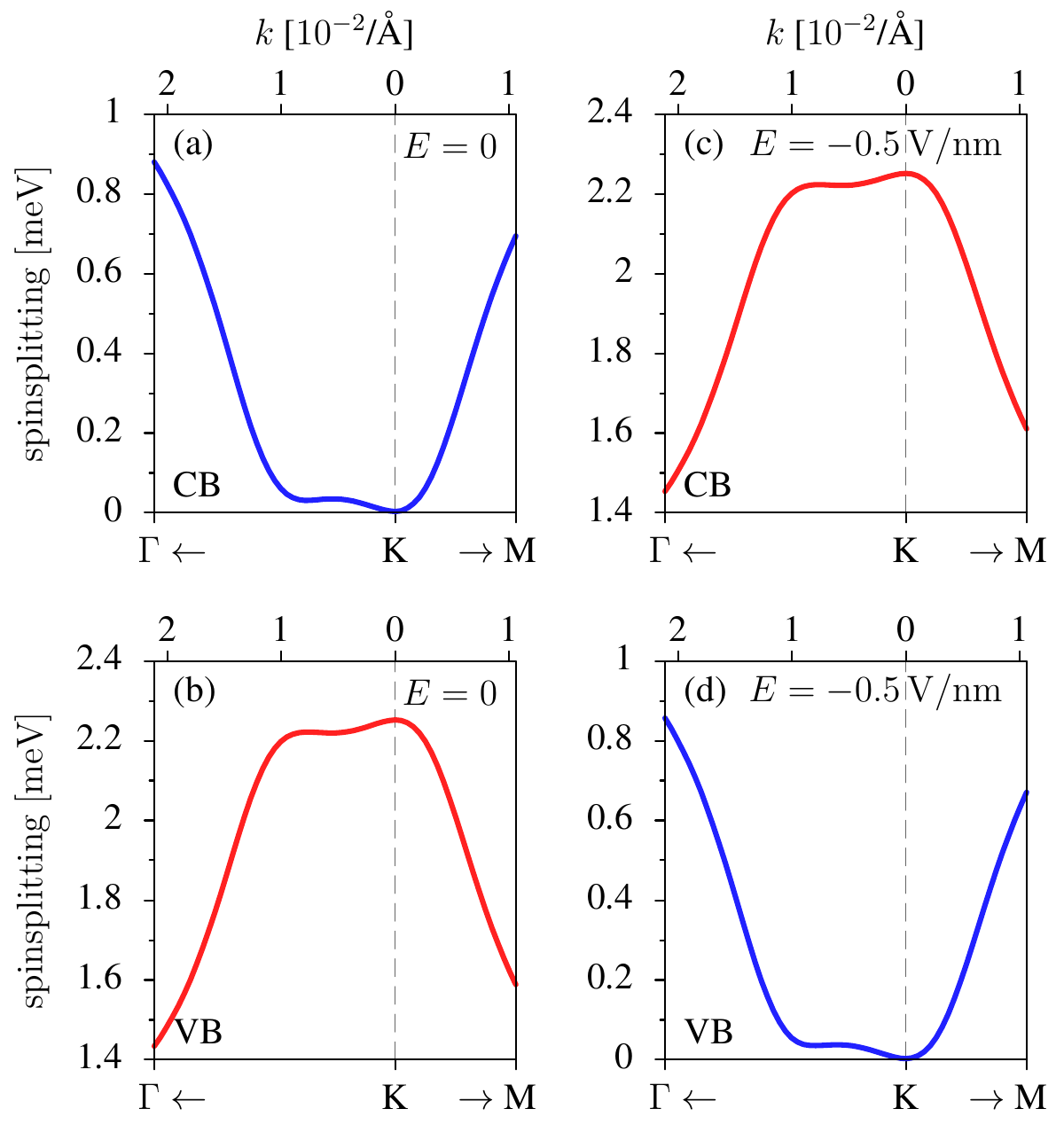}
	\caption{Calculated spin splitting for low energy bands of bilayer graphene on WSe$_2$ near the K valley for
		(a)~conduction band (CB) at zero electric field,
		(b)~valence band (VB) at zero field,
		(c)~same as in (a) and (d)~same as in (b) but for electric field of -0.5~V/nm.
	}\label{Fig:splitting}
\end{figure}

Across the BLG/WSe$_2$ heterostructure induces in simulated
cell of 2.12~nm$^3$ the dipole of 0.7~Debye with built-in electric
field of 0.267~V/nm. Dependence of dipole as a function of the
applied electric field (we note it is external field not the
displacement field) is linear, see Fig.~\ref{Fig:dipole}(a).
The dipole is compensated by the negative field of about -0.3~V/nm.
Applied electric field influences also energy offset of the
BLG low energy states within the WSe$_2$ bandgap.
In Fig.~\ref{Fig:dipole}(b) we show energy offset of the 
BLG valence band at the K point and the 
valence band maximum of the WSe$_2$. 
For field below -1~V/nm the valence band maximum of the WSe$_2$
is shifted above the BLG valence top and BLG is electron doped.
We note that positive electric field points from WSe$_2$ towards 
BLG.

Spin splitting of the low energy conduction and valence bands
of BLG for the zero and -0.5~V/nm applied electric 
field are shown in Fig.~\ref{Fig:splitting}. Comparing the 
spin splittings within conduction or valence band manifolds, 
the change in spin splitting around K valley is significant. 
The minimal values of the spin-orbit splittings are underestimated
by Quantum ESPRESSO, as discussed in the paper.

\end{document}